\documentclass[pra,twocolumn,showpacs,amsmath,amssymb,superscriptaddress,floatfix]{revtex4-2}
\usepackage{amsmath}
\usepackage{xcolor}
\usepackage{graphicx}
\usepackage{float}
\usepackage{balance}
\usepackage{bbm}
\usepackage[colorlinks=true,allcolors=blue]{hyperref}

\newcommand{\prodal}[2]{\underset{#1}{\overset{#2}{\prod}}}
\newcommand{\sumal}[2]{\underset{#1}{\overset{#2}{\sum}}}
\newcommand{\nn}{\nonumber}
\newcommand{\prob}{\rho}

\begin{document}

% Use the \preprint command to place your local institutional report
% number in the upper righthand corner of the title page in preprint mode.
% Multiple \preprint commands are allowed.
% Use the 'preprintnumbers' class option to override journal defaults
% to display numbers if necessary
%\preprint{}

%Title of paper
\title{Digital Herd Immunity and COVID-19}

% repeat the \author .. \affiliation  etc. as needed
% \email, \thanks, \homepage, \altaffiliation all apply to the current
% author. Explanatory text should go in the []'s, actual e-mail
% address or url should go in the {}'s for \email and \homepage.
% Please use the appropriate macro foreach each type of information

% \affiliation command applies to all authors since the last
% \affiliation command. The \affiliation command should follow the
% other information
% \affiliation can be followed by \email, \homepage, \thanks as well.
\author{Vir B. Bulchandani}
\affiliation{Department of Physics, University of California, Berkeley, Berkeley, California 94720, USA}
\affiliation{Princeton Center for Theoretical Science,  Princeton University, Princeton, New Jersey 08544, USA}
\author{Saumya Shivam}
\affiliation{Department of Physics, Princeton University, Princeton, New Jersey 08544, USA}
\author{Sanjay Moudgalya}
\affiliation{Department of Physics, Princeton University, Princeton, New Jersey 08544, USA}
\affiliation{Department of Physics and Institute for Quantum Information and Matter,
California Institute of Technology, Pasadena, California 91125, USA}
\affiliation{Walter Burke Institute for Theoretical Physics, California Institute of Technology, Pasadena, California 91125, USA}
\author{S. L. Sondhi}
\affiliation{Department of Physics, Princeton University, Princeton, New Jersey 08544, USA}
%\email[]{Your e-mail address}
%\homepage[]{Your web page}
%\thanks{}
%\altaffiliation{}

%Collaboration name if desired (requires use of superscriptaddress
%option in \documentclass). \noaffiliation is required (may also be
%used with the \author command).
%\collaboration can be followed by \email, \homepage, \thanks as well.
%\collaboration{}
%\noaffiliation

\date{\today}

\begin{abstract}
A population can be immune to epidemics even if not all of its individual members are immune to the disease, so long as sufficiently many are immune---this is the traditional notion of herd immunity. In the smartphone era a population can be immune to epidemics {\it even if not a single one of its members is immune to the disease}---a notion we call ``digital herd immunity'', which is similarly an emergent characteristic of the population. This immunity arises because contact-tracing protocols based on smartphone capabilities can lead to highly efficient quarantining of infected population members and thus the extinguishing of nascent epidemics. When the disease characteristics are favorable and smartphone usage is high enough, the population is in this immune phase. As usage decreases there is a novel ``contact-tracing phase transition'' to an epidemic phase. We present and study a simple branching-process model for COVID-19 and show that digital immunity is possible regardless of the proportion of non-symptomatic transmission. 
% Our analysis yields new theoretical tools for quantifying how wide and deep a tracing-and-testing network must be to prevent epidemics with substantial non-symptomatic spreading.
% We believe this is a promising strategy for dealing with COVID-19 in many countries such as India, whose challenges of scale motivated us to undertake this study. 
% in the first place and whose case we discuss briefly.
\end{abstract}

% insert suggested keywords - APS authors don't need to do this
%\keywords{}

%\maketitle must follow title, authors, abstract, and keywords
\maketitle

% body of paper here - Use proper section commands
% References should be done using the \cite, \ref, and \label commands
\section{Introduction}
Recent events have challenged the public health infrastructure worldwide for controlling the spread of contagious diseases. This difficulty is partly due to the novel pathogen involved and partly due to some unusual characteristics of COVID-19~\cite{jhu_intro}. Specifically, the infection appears to be transmitted through a large number of asymptomatic and pre-symptomatic cases~\cite{jhu_intro,wei2020presymptom}.

%  With the progression of time, it has become clear that until enough people have been vaccinated, masks and proper indoor air filtration and ventilation~\cite{Howarde2014564118,Lipinksi2020ventilation} are major sources of intervention. However, looking back, with limited information in the initial stages of the pandemic, two main approaches to controlling the exponential growth in the number of infected people were employed. The first is continuous monitoring of entire populations via regular testing, which can identify new infections already during their latent phase and thus end non-symptomatic transmission. The second is the established method of ``contact tracing''~\cite{Fraser}, in which people who have been exposed to newly identified infected people are isolated before they have a chance to infect others. In principle, either approach is capable of ending the epidemic, and are the first lines of attack when detailed information about the transmission of a disease is not known, which is why we restrict our discussion to these interventions.
 
% Population-level testing can be expensive and can involve delays. Traditional contact tracing, done by teams of health officials relying on interviews with newly identified cases, is also not up to the task today, as it fails if non-symptomatic transmission is too frequent~\cite{Fraser, Eames}.  %We note that the possibility of using mobile networks to study disease dynamics has been explored in the past~\cite{Salathe,Yoneki,Gartner}.

In the early days of the pandemic, few public health interventions were available; access to testing was limited, there was no vaccine and mechanisms for transmission of the disease were poorly understood. Faced with this situation, most governments chose to restrict their populations' movements through the introduction of social distancing measures. Some countries complemented this approach with the established method of ``contact tracing''~\cite{Fraser}, in which individuals who have been exposed to newly identified infections are isolated before they have a chance to infect others. Both interventions have proved to be effective for reducing the effective reproduction number $R$. At the time of writing, it is widely appreciated that multiple ``layers of protection'' serve to diminish $R$; interventions adopted more recently include vaccines~\cite{astrazeneca,moderna,pfizer} (though these are still far from ubiquitous) and technological improvements to face masks and indoor air filtration~\cite{Howarde2014564118,Lipinksi2020ventilation}.

This paper is motivated by an urgent practical question that arose at the start of the COVID-19 pandemic: given a disease with a high rate of non-symptomatic transmission, is contact tracing a useful intervention?  One reason for focusing on contact tracing is that it is among the first lines of defence when faced with a new contagious disease; it requires little research overhead compared to pathogen-specific interventions such as vaccines. As such, this question has practical relevance well beyond the specific context of COVID-19~\cite{Fraser}.

Conventional wisdom dictates that traditional contact tracing, executed by teams of health officials who interview newly infected individuals, fails if non-symptomatic transmission is too frequent~\cite{Fraser, Eames}. Fortunately, we live in the smartphone era, and it has been noted that using these devices to record contacts can make the task of tracing them entirely solvable by automating it. This idea has been spelled out in a series of papers~\cite{ferretti2020quantifying,faggian2020proximity} (see also \cite{Salathe,Yoneki,Gartner}) and is the basis for a large set of contact-tracing apps and an Exposure Notifications System developed by Apple and Google ~\cite{googleappleENS}. It is reasonably clear that a perfect digital contact-tracing network will halt a nascent epidemic. The point of this paper is to quantify how far an \emph{imperfect} contact-tracing network can hinder an epidemic with substantial non-symptomatic transmission. We contribute to this emerging field of infectious disease control along two axes.

First, we present a simple model of the early stages of the spread of COVID-19, which allows us to obtain analytical estimates, as a function of a varying amount of non-symptomatic transmission, of the fraction of the population that needs to participate in a digital contact-sharing network in order to prevent new epidemics.
% Our interest in this question was seeded by the very practical question of whether a country like India could use this technology to achieve epidemic control in the initial stages. Indeed, India launched on this enterprise~\cite{indiaapp} and resulted in a large number of downloads of the application. 
While modeling with various differences from our own became available while we were working on this problem~\cite{ferretti2020quantifying,hellewell2020feasibility,braithwaite2020tracingreview}, we feel that our approach has the virtue of making the existence and values of the estimated compliance thresholds transparent. Our estimates for the fraction of the population that needs to own a contact-tracing app to avert a COVID-19 epidemic range from $75\%-95\%$ for $R_0=3$, depending on the fraction of asymptomatic transmission, $\theta=20\%-50\%$, that takes place. For smaller values of $R_0$ due to social distancing this fraction is lower.

% For smaller $R_0$ produced by social distancing this fraction is lower, e.g. $50\%$ for $R_0=2$ and $\theta=0.35$\cite{CDC}. CHECK

Our second contribution is to frame the overall discussion in a language more familiar to physicists and students of complex phenomena more generally---that of phases, phase transitions and emergent properties. The bottom line here is the idea that the immunity of a population to epidemic growth is an emergent, or collective, property of the population. For traditional vaccination or epidemic-induced ``herd immunity'', this feature gets conflated with the fact that individuals can be immune to the disease at issue. But mass digital contact tracing now makes it possible for the population to be immune to epidemic growth {\it even as no individual has immunity to the underlying disease}. We propose to refer to this as the existence of a ``digital herd immunity''. This fits well into the general idea of an emergent property, which does not exist at the level of the microscopic constituents but exists for the collective~\cite{MoreDiff,moon2017emergence}. We would be remiss if we did not note that epidemiologists have previously referred to this state of affairs as ``herd protection'' and ``sustained epidemic control''~\cite{ferretti2020quantifying}. Our intention with the proposed terminology is both to frame a public health goal by including the word digital and to emphasize the emergent nature of a herd immunity. 

Our application of ideas from statistical physics to the theory of contact tracing yields three main dividends compared to earlier approaches. First, we are able to analytically quantify the effectiveness of contact tracing for any recursive depth $n$, from the limit of traditional, manual contact tracing, for which $n=1$, to perfect digital tracing, for which $n=\infty$ in principle. Describing this crossover, which was beyond existing theoretical techniques~\cite{Fraser,Klinkenberg,muller2000contact}, allows us to sharply formulate an important, general principle of disease control: for any proportion of non-symptomatic transmission, tracing and isolation based on a sufficiently widespread contact-tracing network, of sufficient recursive depth, can prevent epidemic spread. The second insight gained through our approach is a point of principle that has been neglected in the public discourse surrounding this issue, namely that once effective contact-tracing protocols are in place, nascent epidemics are extinguished with a probability near one~\footnote{More precisely, the statistical physics of the epidemic-to-immune phase transition implies that in the limit of an infinite population, epidemics are suppressed almost surely beyond a certain threshold population fraction $\phi>\phi_c$ on the contact-tracing network and sufficiently large recursive depth $n$.}. Third, our study of the universal properties of the contact-tracing phase transition allows us to capture the full probability distribution for epidemic sizes as the critical threshold for epidemic control is approached. Our analysis suggests that the contact-tracing phase boundary is not even visible at the resolution of influential previous studies for COVID-19~\cite{hellewell2020feasibility}, that attempted to estimate the critical threshold for contact tracing by numerically simulating the epidemic-size distribution function.

In the balance of this paper we do the following. We begin by summarizing the case for contact tracing as an effective strategy for combating COVID-19. We then introduce a simple branching-process model for the spread of this disease, that incorporates the key features of asymptomatic transmission, pre-symptomatic transmission and recursive contact tracing. We find that there is {\it always} a critical fraction $0 \leq \phi_c <1$ of app ownership, such that take-up of contact-tracing apps by a fraction $\phi > \phi_c$ of the population is sufficient to prevent epidemic spread. We provide an analytical formula for this threshold, which is verified against detailed numerical simulations, and characterize the universal features of the resulting ``contact-tracing phase transition''. In the final discussion, we turn to practical matters that arise in real-world implementations of digital contact tracing, such as the need for population-wide testing.

\section{A Model for App-Based Contact Tracing}
\subsection{Motivation and relevance to COVID-19}

Traditional contact tracing is a multi-stage process. First, one identifies symptomatic, infected individuals. Next, one finds the people they came into close contact with during their infectious period. Finally, one treats or isolates these people before they can go on to infect others. Manual contact tracing becomes difficult for infections that have a period before the onset of symptoms when an exposed person is contagious (the $\Omega$ period). Further delay in finding the symptomatic person and their contacts could lead to tertiary infections, making it difficult to control an outbreak. For COVID-19, the incubation period is thought to be around 5-6 days, while the $\Omega$ period is estimated to be 1-3 days \cite{wei2020presymptom,nishiura2020serial}. The time before becoming contagious, or the latent period $L$, is around 4 days. Stochasticity of these times aside, it is reasonable to expect that if $\Omega<L$ on average, and if the exposed contacts of an individual can be traced before they become infectious, then an epidemic could be prevented. However, the delays typical for manual contact tracing, even just one or two days, can render contact tracing completely ineffective for COVID-19, given the typical $L$ and $\Omega$ periods; this conclusion is supported by detailed numerical simulations~\cite{hellewell2020feasibility, ferretti2020quantifying}. 

This is where digital contact tracing comes in. A smartphone application could enable instant isolation of an infected person and their network of contacts. This halts the transmission chain, because infected contacts cannot infect others during their latent period. The question immediately arises of how widespread such tracing needs to be in order to prevent an epidemic, and this question is the focus of our paper. Below we present a simple model that captures the essential features of disease spread necessary to tackle this problem. 

To place our work in context, the classic quantitative analyses of the efficacy of contact tracing~\cite{Fraser,Eames}, from before the smartphone era, showed that traditional, manual contact-tracing protocols become useless when the rate of non-symptomatic spreading, $\theta$, is too high. By contrast, app-based approaches allow for ``recursive'' contact tracing, whereby contacts of contacts can be traced to an arbitrary recursive depth, at no additional cost. The effectiveness of recursive contact tracing has been studied in previous work; mathematically rigorous results exist in simple limits~\cite{muller2000contact,Mueller2} and detailed numerical simulations have been performed in analytically inaccessible regimes~\cite{Klinkenberg}. Some recent works have provided quantitative estimates for the effectiveness of non-recursive contact-tracing in the specific context of COVID-19~\cite{ferretti2020quantifying,hellewell2020feasibility,faggian2020proximity}. Our results should be viewed as complementary to these studies. One advantage of the model that we propose is its simplicity; this allows for a more thorough analytical understanding of the contact-tracing phase transition than in previous works, for \emph{any} recursive depth, $0 \leq n \leq \infty$.

\subsection{A branching-process model}

\begin{figure}[t]
    \centering
    \includegraphics[trim=0cm 2cm 0cm 7cm,width=0.95\linewidth]{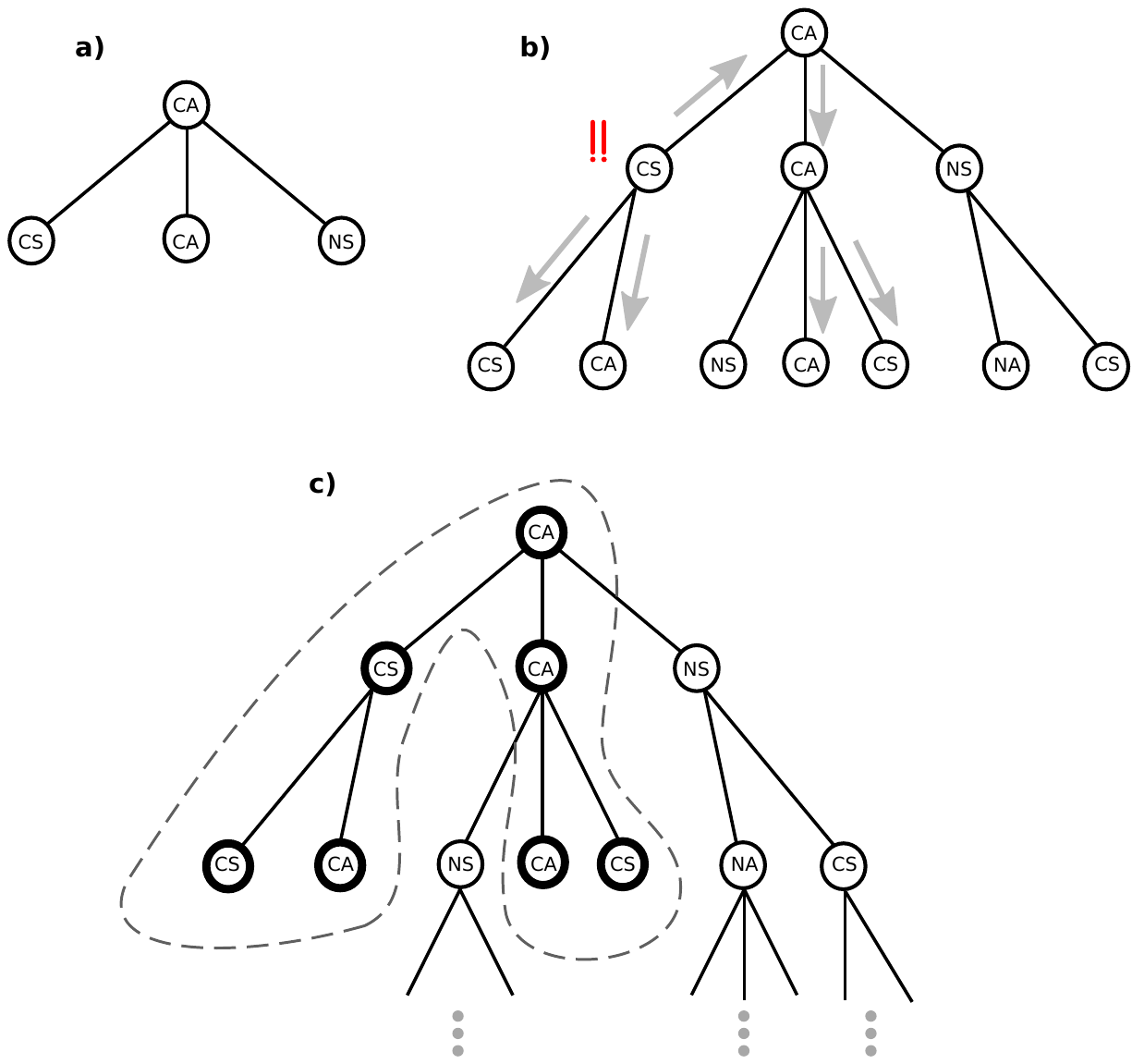}
    \caption{An illustrative realization of our branching-process model. \textit{a)} An in-network asymptomatic individual (CA) infects $R_0=3$ people, whose category of infection is chosen independently and uniformly subject to parameters $\theta$, the fraction of the population that presents asymptomatic cases, and $\phi$, the fraction of the population using a contact-tracing app. \textit{b)} A CS infection triggers an alert on the contact network, but the CS individual and everybody else in their generation is still able to infect people before the contact network is triggered (while we depict $R_S=2$, we also simulate more realistic values $R_S=0,\,1$). The arrows show the alerts sent to everyone connected to the CS individual by the contact network. \textit{c)} Once the alert is sent out, everybody in the contact-network-connected component of the CS individual can be quarantined immediately (thick circles) without giving rise to further disease spread, since they are in the latent (non-contagious) periods of their infections. Meanwhile, every individual off the contact network continues to transmit the disease freely. The ``recursive'' aspect of such contact tracing corresponds to the arrow going back in time in Fig. (1b); the middle branch of infections would be missed by a traditional, non-recursive approach.}
    \label{Fig1}
\end{figure}

Suppose we have an epidemic spreading through an infinite population of susceptibles, in discrete time, and infecting $R^t$ members of the population at each time step $t =0,1,2,\ldots$ (here, $R$ is an ``effective reproduction number'' that depends on the detailed properties of the epidemic spread, including the basic reproduction number $R_0$).
This is a generic model for a spreading epidemic at short times. The total number of infections $I^{\mathrm{tot}}$ scales as
\begin{equation}
I^{\mathrm{tot}} \propto \begin{cases} \frac{1}{1-R} & R < 1
\\ \infty & R \geq 1, \end{cases}
\end{equation}
and if $R<1$, the epidemic has been controlled.

We want to understand which $R$ best captures the effect of mobile-phone-based contact tracing. From a statistical physics perspective, $R$ is the single relevant parameter controlling the epidemic spread, and drives a phase transition from an ``epidemic phase'' to an ``immune phase'' as $R$ decreases below $R=1$, which we shall elaborate on below. For the purposes of modelling epidemic spread, the key question is which ``microscopic'' degrees of freedom must be included to obtain a realistic estimate for $R$.

To this end, we consider three parameters that implicitly determine $R$ : the fraction of the population that will present asymptomatic cases ($\theta$), the fraction of the population using a contact-tracing application ($\phi$), and the basic reproduction number for an individual who eventually shows symptoms ($R_S$).
$R_S$ is a combined measure of the number of pre-symptomatic infections and the efficacy of quarantine: in the limit of perfect isolation after showing symptoms, $R_S$ is precisely the number of people that a symptomatic individual infects during their $\Omega$ period, as defined above. We assume that $R_S$ is independent of whether a symptomatic individual is on the contact-tracing network or not.

The effects of these parameters on the growth of the epidemic (or $R$) are studied using a simple branching-process model, where all infectious individuals are either symptomatic (S) or asymptomatic (A), and either on the app-based contact tracing network (C) or not (N). In an an uncontrolled setting, all types of infectious cases are assumed to proliferate with $R_0=3$, which is a reasonable estimate \cite{kucharski2020early} for COVID-19 \footnote{There are also some higher estimates in the literature, e.g. based on the early epidemic dynamics in Wuhan \cite{Sanche2020.02.07.20021154}. Given the greatly increased state of awareness of the disease at this point it seems reasonable to assume that such higher values are not relevant to disease dynamics today.}. Suppose that the outbreak starts from a single infected individual at time $t=0$ (``Patient Zero''). In our discrete time (generational) model, Patient Zero infects $R_0$ other people at time $t=1$, and each new infection is assigned to one of the categories $\{CA, CS, NA, NS\}$ randomly, with probabilities that are determined by the values of $\theta$ and $\phi$.
Individuals infected at the beginning of each generation are assumed not to infect anyone else after that generation has elapsed.
Whenever a symptomatic individual on the contact network (CS) is encountered during this branching process, the contact network is triggered, and all people connected to the CS individual by the network, through either past or present infections, are placed in quarantine. As discussed earlier, since pre-symptomatic infections are common for COVID-19, our model includes the possibility that a CS individual infects $R_S$ people by the time they trigger the contact network. As a consequence, non-CS individuals in the same generation are also allowed to infect the next generation before the activation of the contact network (see Fig.~1c).
A few time-steps of the model are illustrated explicitly in Fig. \ref{Fig1}, together with the implementation of recursive contact tracing via removing connected components of the contact graph.
Different combinations of the parameters $\theta$, $\phi$ and $R_S$ lead to an effective reproduction number $R$ distinct from the bare reproduction number $R_0$, and we expect epidemic growth to be suppressed whenever $R<1$.

\begin{figure}[t]
    \centering
    \includegraphics[width=0.99\linewidth]{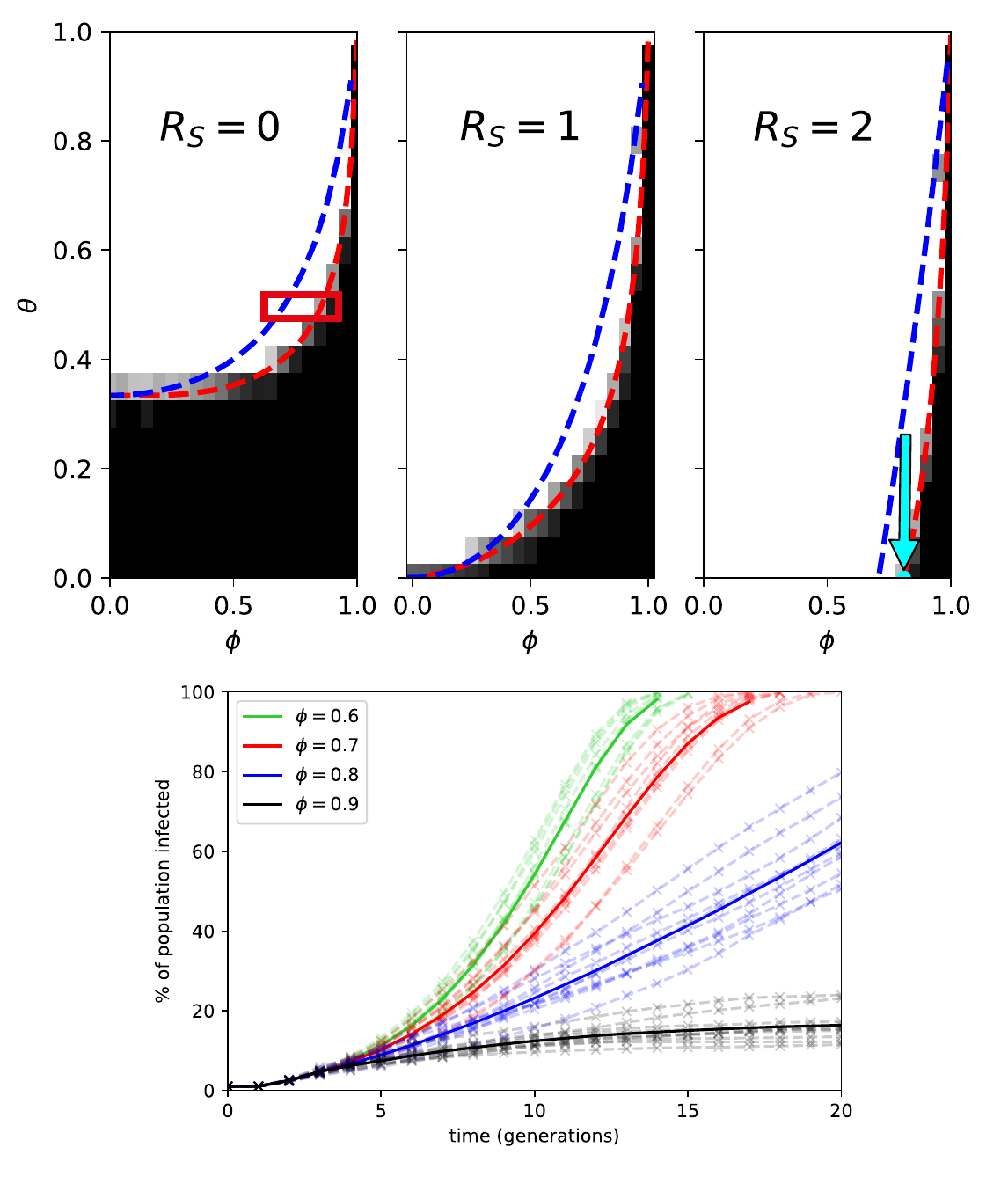}
    \caption{\textit{Top:} Phase diagrams of epidemic control for $R_0=3$, where the tuning parameters are $\theta$, the rate of asymptomatic transition, and $\phi$, the fraction of contact-tracing app ownership among the population. $R_S$ denotes the basic reproduction number for pre-symptomatic transmission. Each phase diagram was generated from $4000$ microscopic simulations of $20$ generations of disease evolution on $10,000$ nodes, of which $100$ nodes were initially infected at random. A black square denotes epidemic control (average growth in the cumulative number of infections over the generations $16-20$ is less than $0.25\%$ of total population). This is grayscaled continuously to white for late-time growth exceeding $2.5\%$ per generation or full epidemic spread before $20$ generations have elapsed. Dashed red curves show exact results for $10$-step contact tracing, while dashed blue curves show an easy-to-use approximation to the exact result. Both formulae are presented in Sec.~\ref{sec:Analytical}. The exact critical point for $\theta=0$ and $R_S=2$, derived in Appendix \ref{AppendixA}, is marked by a cyan arrow. \textit{Bottom:} Sample simulations from the encircled region in the $R_S=0$ phase diagram. Curves (solid) denote cumulative number of infections as a percentage of total population, averaged over 10 samples (dashed), with $\theta=0.5$ fixed and $\phi$ varied from $\phi=0.6$ to $\phi=0.9$.}
    \label{Fig2}
\end{figure}

Numerical simulations of this model were performed on $10,000$ nodes with $100$ initial infections, without replacement; the results are summarized in Fig. \ref{Fig2}. The location of the phase boundaries were verified to be independent of both doubling the system size and doubling the number of samples averaged per point shown on the phase diagram, to within the resolution of the phase diagram. Our numerics are consistent with the hypothesis that for any given fraction of asymptomatic transmission $0 \leq \theta < 1$, and any presymptomatic reproduction number $0 \leq R_S \leq R_0$, there is a critical point $\phi_c(\theta)$, corresponding to the onset of ``digital herd immunity'': epidemic control occurs for a fraction of app owners $\phi_c(\theta)<\phi \leq 1$. For realistic COVID-19 parameter values, $R_0=3$, $R_S=1$ and $\theta = 0.2-0.5$ ~\cite{kucharski2020early,ferretti2020quantifying}, we find that $\phi_c(\theta)=75\%-95\%$, illustrating that when both presymptomatic and asymptomatic transmission are taken into account, the rate of app coverage necessary to prevent an epidemic can be rather high. Some practical implications of this point are raised in the final discussion.
\subsection{The contact-tracing phase transition}\label{sec:Analytical}
We now describe the sense in which our branching-process model exhibits a phase transition. Consider the disease dynamics seeded by a single initial infection, Patient Zero, at $t=0$. As the disease spreads, there are two possibilities: either the epidemic seeded by Patient Zero terminates at some finite time, or it continues to spread indefinitely. In branching process theory \cite{GrimmettProb}, this dichotomy is captured by the ``probability of ultimate extinction'', $q$, which is the probability that the epidemic seeded by Patient Zero terminates at some $t<\infty$. 

To make the connection with statistical physics, consider the quantity $\rho = 1-q$, which is the probability that the epidemic seeded by Patient Zero spreads for all time. This defines an order parameter for the epidemic-to-immune phase transition, in the following sense. If $\rho>0$, an epidemic can spread with non-zero probability, and the population is in an ``epidemic phase''. If $\rho=0$, epidemics are almost surely contained, and the population is in an ``immune phase''. In fact, $\rho$ is precisely the order parameter for a site percolation phase transition~\footnote{To see this, first designate all nodes of this tree ``empty'', except for the root. Next, embed the branching processes model in this tree in the natural way, with Patient Zero at the root, and ``occupy'' nodes according to their probability of infection. The the probability of formation of an infinite cluster of infections, $\rho$, is precisely the probability that occupied sites percolate on the Cayley tree, which is usually taken as the order parameter in percolation theory~\cite{Grimmett_1999}. We note that epidemic-to-immune phase transitions are often interpreted in the language of bond percolation~\cite{Newman}. However, on the Cayley tree the bond and site percolation thresholds are equal; since the site percolation picture is slightly more natural from the branching process viewpoint, we work with sites rather than bonds throughout this paper.} on the infinite, rooted, Cayley tree with bulk coordination number $z=1+R_0$.

We argued above that the epidemic-to-immune phase transition is driven by a single relevant parameter, the effective reproduction number, $R$. Let us now make this statement precise. For simple epidemic models, the underlying branching process is Markovian, and we can define $R$ to be the mean number of new infections generated by an infected node. It is then a rigorous result that $\rho=0$ for $R \leq 1$ and $\rho>0$ otherwise~\cite{GrimmettProb}. For such models, the connection between epidemic spread and percolation transitions has been known for some time~\cite{Grassberger1982,Cardy_1983,Cardy_1985,Newman,Kenah}. By contrast, for the model studied in this paper, the possibility of tracing successive contacts means that the disease dynamics is no longer Markovian; there are correlations between generations that preclude a simple definition of $R$, and earlier theoretical results do not apply. For the same reason, our model has no simple interpretation in terms of site or bond percolation once contact tracing is included (i.e. for $\phi > 0$ and $n \geq 1$).

The main technical innovation in our work is surmounting this breakdown of the Markov property: we develop generating function methods that allow for \textit{exact} summation of non-Markovian contact-tracing processes, to any desired order. We show that despite inter-generational correlations, the critical behaviour is determined by a function $R_n(\phi,\theta)$, which can be viewed as a ``mean number of new infections'', suitably averaged over time. In particular, $R_n(\phi,\theta)$ controls the ultimate fate of the epidemic, and the critical line for $n$-step contact tracing is given by an implicit equation
\begin{align}
\label{ECP}
R_n(\phi,\theta) = 1
\end{align}
in $\phi$ and $\theta$; details of the calculation and a full expression for $R_n(\phi,\theta)$ are presented in the Supplementary Material. Taking the limit as $n\to \infty$ yields the exact critical line for contact tracing to arbitrary recursive depth; however, for any $n\geq 1$, this does not seem to be expressible in closed form, except at its endpoints. Fortunately, the function $R_n$ is found to converge rapidly in its arguments with increasing $n$. Fig. \ref{Fig2} depicts results from Eq. \eqref{ECP} with ten-step contact tracing, $n=10$, and shows excellent agreement with stochastic numerical simulations. We note that the values obtained for $n=10$ are already well-converged, in the sense that they do not vary significantly for $n>10$. The sensitivity of the efficacy of contact tracing to tracing depth is addressed in more detail in a companion paper~\cite{shivam2021recursive}.

We now discuss the universal properties of this phase transition. For concreteness, let us parameterize the critical line as $(\phi_c(\theta),\theta)$. For fixed $\theta$ in the domain of $\phi_c$, a transition from an epidemic to an immune phase occurs as $\phi \to \phi_c(\theta)^-$. We call this transition the ``contact-tracing phase transition'', because it is controlled by the population fraction on the contact-tracing network. To capture the universal properties of this transition, it is helpful to consider the random variable $|C|$, which is the size of the infected cluster seeded by Patient Zero in a single realization of the branching process. On the immune side of the transition, $|C|$ is almost surely finite, and the risk of epidemics is captured by the mean cluster size, $\mathbb{E}_{\phi,\theta}(|C|)$. On the epidemic side of the transition, the mean cluster size diverges, and the order parameter $\rho_{\phi,\theta} = \mathbb{P}_{\phi,\theta}(|C| = \infty)$ better quantifies the risk of epidemic spread. In the vicinity of the critical point $\phi=\phi_c(\theta)$, both quantities are characterized by universal critical exponents,
\begin{align}
\rho_{\phi,\theta} &\sim (\phi_c(\theta)-\phi)^{\beta}, \quad \phi \to \phi_c(\theta)^-, \\
\mathbb{E}_{\phi,\theta}(|C|) &\sim \frac{1}{(\phi-\phi_c(\theta))^\gamma}, \quad \phi \to \phi_c(\theta)^+.
\end{align}
In the Supplementary Material, we show that $\gamma=\beta=1$ for any finite $n$, demonstrating that the contact-tracing phase transition lies in the universality class of mean-field site percolation~\cite{Grimmett_1999}. The scaling theory of percolation transitions then implies a universal scaling form for the distribution of epidemic sizes,
\begin{equation}
\mathbb{P}_{\phi,\theta}(|C|=N) \sim N^{-3/2}f(N/N_\xi), \quad \phi \to \phi_c(\theta),
\end{equation}
where $f$ is a scaling function with exponentially decaying tails, and the cluster correlation length $N_{\xi} \sim (\phi-\phi_c(\theta))^{-2}$ as $\phi \to \phi_c(\theta)$.

\begin{figure}[t]
    \centering
    \includegraphics[width=0.99\linewidth]{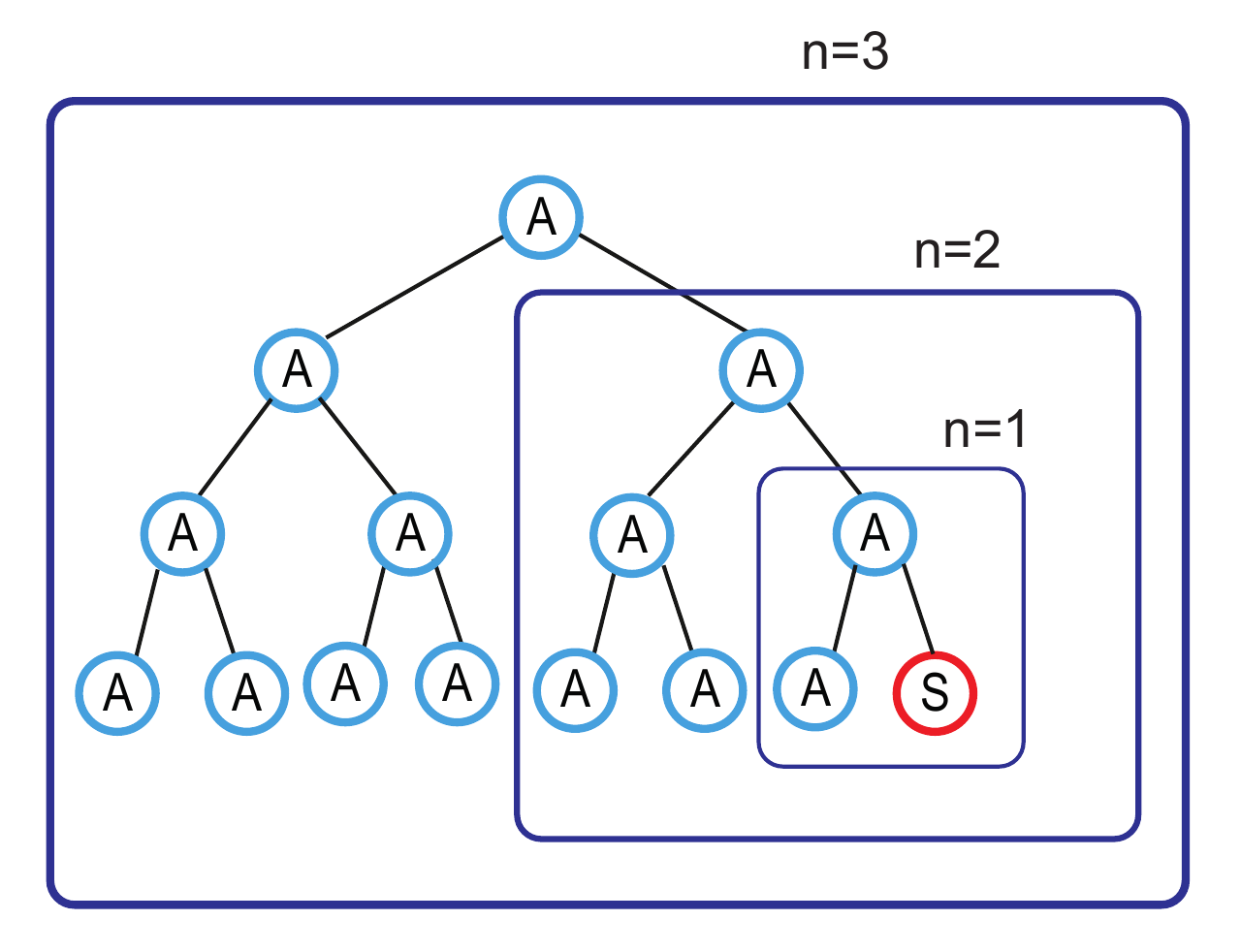}
    \caption{The non-locality of the branching-process dynamics increases with the contact-tracing depth, $n$. The figure depicts a growing cluster of asymptomatic infections that are all on the contact network. As the recursion depth $n$ for contact tracing increases, a single symptomatic infection allows for isolation of increasingly large asymptomatic clusters, up to a size that grows exponentially in $n$. In the limit $n\to \infty$, this results in a discontinuous phase transition at $(\phi,\theta) = (1,1)$; see Appendix \ref{AppendixC} for details.} 
    \label{fig:recursion}
\end{figure}

As the recursive tracing depth $n \to \infty$, we find that such mean-field behaviour breaks down at the endpoint of the critical line $(\phi, \theta) = (1,1)$, which exhibits a \textit{discontinuous} phase transition, reflecting the non-locality of the underlying branching process; see Fig. \ref{fig:recursion}. The emergence of a discontinuous percolation transition on the Bethe lattice is highly unusual, and suggests that the non-local character of the $n=\infty$ contact-tracing transition fundamentally distinguishes it from the percolation-type phase transitions that have arisen in related settings~\cite{Grassberger1982,Cardy_1983, Cardy_1985,Newman,Kenah, DrosselSchwabl,DROSSEL1993183}.

Although Eq. \eqref{ECP} is exact, the full expression for $R_n$ is a little cumbersome to rapidly adapt and use. We therefore present a simpler, approximate formula for $R = R_\infty(\phi,\theta)$, based on linear interpolation and a ``mean-field'' assumption, whereby inter-generational correlations are neglected. The resulting approximation to $R$, derived in Appendix \ref{AppendixD}, is given by
\begin{equation}
\label{ApproxRMain}
R \approx R_S (1 - \theta)(1-\phi^2) + R_0 \theta\left(1 - \left(1 - \frac{1}{R_0}\right) \phi^2\right).
\end{equation}
In Fig. \ref{Fig2}, the critical threshold $R=1$ predicted by Eq. \eqref{ApproxRMain} is compared to the exact result, Eq. \eqref{ECP}, for $10$-step contact tracing, and shows reasonably good quantitative agreement.

\section{Discussion}

We have introduced a simple branching-process model for early-stage epidemic spread, which both retains a degree of analytical and numerical tractability and is sufficiently expressive to model complicated features of COVID-19 spreading and control, for example pre-symptomatic transmission, as distinct from asymptomatic transmission, and recursive contact tracing. Using this model, we have obtained predictions for the app take-up fraction needed to provide digital herd immunity as a function of $R_0$ and the asymptomatic transmission frequency $\theta$. Our conclusions are significant beyond the immediate context of COVID-19: this paper can be read as a theoretical demonstration that digital contact tracing is effective even when the rate of non-symptomatic transmission is high and not everybody is on the contact network. This was not at all clear, even as a matter of principle, at the start of the COVID-19 pandemic.

Subsequent developments, in countries such as South Korea, Vietnam, Japan and Taiwan, have confirmed that digital contact tracing can be deployed successfully against COVID-19~\cite{lewis2020many}. India's digital contact tracing effort, which provided the initial stimulus for our study, was also substantial~\cite{indiaapp}, with over 100 million individual downloads of the mobile application. However, despite such successes, serious adoption of digital contact tracing during the present crisis has been rather limited. This is surprising, given its basic importance as a tool for epidemic prevention~\cite{Fraser}. We now examine some of the reasons behind this neglect.

One concern, which has severely hindered the widespread adoption of digital contact-tracing technology in Europe and the USA, is protecting the privacy of individuals on the contact tracing network. As a purely technical problem, this is easily surmountable but a lack of public faith in governmental handling of data has no simple remedy. The seriousness of this obstacle depends on the details of the social contract between the state and its population, which tends to be more rigid in countries where digital contact tracing has been successful~\cite{lewis2020many}.

A more tractable problem is that recursive digital contact tracing can rapidly lead to an excessive number of ``alerts'', in the sense of Fig. \ref{Fig1}, unless it is accompanied by a widespread testing program to minimize the number of false positives. Conversely, if widespread testing is already in place, then digital contact tracing becomes a highly effective tool for epidemic prevention. Unfortunately, widespread testing was not in place for many months after the start of the pandemic. Published estimates for the required testing rate~\cite{Jha,Sid} further tended to underestimate the testing overhead due to recursive tracing of asymptomatic infections. A revised estimate, which correctly accounts for both tracing contacts of contacts and asymptomatic infections, is given in Appendix \ref{AppendixE}.

% We now consider the practical applicability of our results to India, whose particular challenges provided the initial stimulus for this work. India's overall smartphone coverage is around 40\%~\cite{trai}. However, this percentage is much higher in the major cities which are the primary challenge and since the start of the epidemic $R_0$ has not gone much higher than $2$. Further almost 90\%~\cite{trai} of Indians use some sort of wireless phone, which can be included on a digital contact tracing network to various degrees using cell tower based triangulation and SMS broadcast. So when compared with the challenges of manual contact tracing, it is clear that the digital route is much more scalable. Indeed, India has had considerable success with this strategy already, with over 100 million downloads and $100,000$ contacts traced.

We end with some comments towards the future. In terms of statistical mechanics, it would be useful to generalize our computations to take real-world heterogeneity of $R_0$ into account~\cite{Lloyd-Smith2005} and to examine the course of epidemics on realistic graphs when $\phi < \phi_c$ (see Ref. \cite{shivam2021recursive} for an extension of the model introduced above to arbitrary contact networks). In terms of infectious disease control we believe digital control has a promising future---it is a non-trivial idea with enormous scalability and can be greatly strengthened by combining it with wearable diagnostics that track the health status of the wearer. Variants of COVID-19 will be with us for the forseeable future and influenza is already endemic.
It seems realistic to aim at greatly reducing the annual incidence of both of these diseases by adding digital control to the existing toolkit. Finally, we believe that one should start to think of the herd immunity of a population as deriving from a combination of natural immunity, vaccination, use of personal protective equipment {\it and} digital immunity.

%One worthwhile extension of our model would be to consider network connectivities beyond the tree structure considered in this work, for example, by simulating the branching-process model on small-world networks. This is of practical importance because such networks are known to better reflect realistic human contact networks~\cite{SmallWorld,EpiComplex}; indeed, network structure has previously been shown to influence the effectiveness of contact tracing~\cite{huerta2002contact,Eames,kiss2005disease,Kiss2005InfectiousDC}. An interesting theoretical question is to understand how far our exact results for the contact-tracing phase transition on the Bethe lattice, including the unusual, discontinuous critical point in the limit $n\to \infty$, extend to this more general setting.

\vspace{2mm}
% Specify following sections are appendices. Use \appendix* if there
% only one appendix.
%\appendix
%\section{}

% If you have acknowledgments, this puts in the proper section head.

\section*{acknowledgments}
We would like to thank the Principal Scientific Adviser to the Government of India, Professor K. VijayRaghavan, for interesting us in this question and for discussions of India's Aarogya Setu contact tracing app, Professor Bryan Grenfell for sharing his wisdom regarding epidemiology at an extremely hectic time and Dr.~Shoibal Chakravarty for continuing discussions on all aspect of India's COVID-19 challenges.

% Create the reference section using BibTeX:
%\balance
\bibliography{ref}
\onecolumngrid
%\balance
\appendix
\section{Critical behaviour along the line $\theta=0$}\label{AppendixA}
In this appendix, we derive the critical behaviour along the line $\theta=0$. 
Since all infections on this line are symptomatic, we denote $R_S \equiv R$.
As we demonstrate below, the exact critical point can be obtained in closed form, and is found to be
\begin{equation}
\phi_{c}  = \frac{R-1 + \sqrt{(R-1)(R+3)}}{2R}, \quad R \geq 1,
\end{equation}
which matches our numerical phase diagram to within the resolution of the plot; see Fig.~\ref{Fig2} for an example with $R=2$.

Let us now summarize some basic facts about the critical behaviour of percolation-type transitions. Recall~\cite{Grimmett_1999} that standard site percolation on the Bethe lattice exhibits five metric-independent critical exponents, which we may denote $\{\alpha,\beta,\gamma,\delta,\Delta\}$. The three scaling relations imply that only two of these are independent. There are three more metric-dependent critical exponents, $\{\nu,\rho,\eta\}$, which can be expressed in terms of the previous exponents using the hyperscaling relations in six dimensions.

We will focus on the two independent exponents $\gamma$ and $\beta$, since these are the easiest to calculate. They control the critical behaviour of the mean cluster size $\mathbb{E}_p(|C|)$ and the percolation probability $\mathbb{P}_p(|C|=\infty)$ (probability of formation of an infinite cluster) respectively.
In terms of the site occupation probability $p$ and its critical value $p_c$, they are defined by 
\begin{equation}
\mathbb{E}_p(|C|) \sim \frac{1}{(p_c-p)^\gamma}, \quad p \to p_c^-,\;\;\;\textrm{and}\;\;\;\mathbb{P}_p(|C|=\infty) \sim (p-p_c)^{\beta}, \quad p \to p_c^+.
\end{equation}
Site percolation on the Bethe lattice with $z \geq 3$ lies in the universality class of mean-field percolation, and exhibits critical exponents $\gamma=\beta=1$~\cite{Grimmett_1999}.
\subsection{Mean cluster size}
To compute the mean cluster size, we adapt a method introduced by Fisher and Essam for counting clusters of a given size in the Bethe lattice~\cite{Fisher}.
We define a probability generating function for the cluster size
\begin{equation}
    B(\phi, x) = \sum_{s = 1}^\infty \mathbb{P}_{\phi}(|C| = s \, | \mathrm{\, initial \, node \, infected}) x^s,\;\; \phi > \phi_c, 
\label{eq:fullgenfunc}
\end{equation}
where $|C|$ denotes the cluster size and $\mathbb{P}_{\phi}(|C| = s \, | \mathrm{\, initial \, node \, infected})$ denotes the probability of obtaining a cluster of size $s$ from an initial infected node.
Since susceptible individuals are on the network ($C$) with probability $\phi$ and off the network ($N$) with probability $(1-\phi)$, $B(\phi, x)$ can be expressed as
\begin{equation}
B(\phi, x) = \phi B_C(\phi,x) + (1-\phi) B_N (\phi, x),\;\;\;B_{C/N}(\phi,x) = \sum_{s=1}^\infty \mathbb{P}_{\phi}(|C| = s \, | \mathrm{\, initial \, node \, infected \, and \,} C/N) x^s.
\label{eq:genfunc}
\end{equation}
Using Eq.~(\ref{eq:genfunc}), the expression for the mean cluster size reads
\begin{equation}
    \mathbb{E}_{\phi}(|C|) = \partial_x B\big{|}_{x = 1} = \phi\ \partial_x B_C\big{|}_{x = 1} + (1 -\phi) \partial_x B_N\big{|}_{x = 1}. 
\label{eq:expectation}
\end{equation}
If a node of type $C$ infects another node of type $C$, the latter cannot transmit infection further. However, a node of type $N$ can infect other nodes freely. This implies the recurrence relations
\begin{align}
B_C &= x\left[\phi x + (1-\phi)B_N\right]^{R},\label{eq:BC}\\
B_N &= x\left[\phi B_C + (1-\phi) B_N\right]^{R} = x B^R\label{eq:BN},
\end{align}
for the coefficients of each probability generating function.
Differentiating at $x=1$, and using the normalization constraint $B_{C/N}(\phi, x = 1) = 1$, we obtain
\begin{align}
    \partial_x B_C\big{|}_{x =1} &= 1 + R \left(\phi + (1-\phi)\ \partial_x B_N\big{|}_{x=1}\right)\nn \\
    \partial_x B_N\big{|}_{x =1} &= 1 + R \left(\phi\ \partial_x B_C\big{|}_{x=1} + (1-\phi)\ \partial_x B_N\big{|}_{x=1}\right).
\end{align}
Solving these linear equations, we find that
\begin{align}
\partial_x B_C\big{|}_{x=1} &= \frac{R^2 \phi^2 - R(R-1) \phi + 1}{R^2 \phi^2 - R(R-1)\phi - (R-1)},\\
\partial_x B_N\big{|}_{x=1} &= \frac{R^2 \phi^2+R\phi+1}{R^2 \phi^2 - R(R-1)\phi - (R-1)}.
\label{eq:partials}
\end{align}
Thus, using Eqs.~(\ref{eq:partials}) and (\ref{eq:expectation}), we obtain the mean cluster size:
\begin{align}
\nonumber \mathbb{E}_{\phi}(|C|) = \frac{R \phi+1}{R^2\phi^2 - R(R-1)\phi - (R-1)}.
\end{align}
Denoting the roots of the denominator by
\begin{equation}
\phi_{\pm}  = \frac{R-1 \pm \sqrt{(R-1)(R+3)}}{2R},
\label{eq:roots}
\end{equation}
we may write
\begin{equation}
\mathbb{E}_{\phi}(|C|) = \frac{R \phi+1}{R^2(\phi-\phi_+)(\phi-\phi_-)}.
\label{eq:mean}
\end{equation}
Since $0 \leq \phi \leq 1$, it follows that the mean cluster size has a simple pole at $\phi = \phi_+$, and assumes a physical, positive value only when $\phi > \phi_+$. Thus, the exact critical point lies at $\phi_c = \phi_+$. (The unphysical, negative value in Eq. \eqref{eq:mean} in the percolating regime $\phi < \phi_c$ reflects the divergence of the mean cluster size due to the infinite cluster. Meaningful results in the percolating regime can be recovered by conditioning on the event $\{|C|<\infty\}$ \cite{Grimmett_1999}, but we will not pursue this here.)

In the vicinity of the critical point $\phi_c=\phi_+$, we obtain
\begin{equation}
\mathbb{E}_{\phi}(|C|) \sim \frac{R\phi_c+1}{R\sqrt{(R-1)(R+3)}} \frac{1}{(\phi-\phi_c)}, \quad \phi \to \phi_c^+,
\end{equation}
from which the critical exponent $\gamma=1$ can be read off.

\subsection{Percolation probability}
To derive $\beta$, we define the probabilities of infinite cluster formation from a source infection that is respectively on or off the contact network:
\begin{equation}
\prob_{C/N}(\phi) = \mathbb{P}_{\phi}(|C| = \infty | \mathrm{\, initial \, node \, infected \, and \,} C/N), \quad \phi \leq \phi_c.
\end{equation}
We now derive recurrence relations to compute $\prob_{C}$ and $\prob_N$.
Since a $C$ node can give rise to an infinite cluster only through infecting an $N$ node, we can obtain an expression for $\prob_C$ in terms of $\prob_N$ as follows. 
Note that $(1 - (1 - \phi) \prob_N)$ is the probability that an infected $N$ node does not lead to an infinite cluster, and hence $(1 - (1 - \phi) \prob_N)^R$ is the probability that none of the $R$ infected nodes lead to infinite clusters.
Thus, the probability that at least one of the nodes infected by an initial $C$ node leads to an infinite cluster is given by
\begin{equation}
    \prob_C = 1 - \left(1 - (1-\phi) \prob_N\right)^R.
\label{eq:Cinfected}
\end{equation}
Similarly, noting that an $N$ node can lead to an infinite cluster via either $C$ or $N$ nodes, the probability that at least one of the infected nodes leads to an infinite cluster is given by
\begin{equation}
    \prob_N = 1-\left[1-(1-\phi)\prob_N - \phi \prob_C\right]^R.
\label{eq:Ninfected}
\end{equation}
Eqs.~(\ref{eq:Cinfected}) and (\ref{eq:Ninfected}) reduce to a single equation for $\prob_N$:
\begin{equation}
\prob_N = 1-\left[1-\phi-(1-\phi)\prob_N + \phi [1-(1-\phi)\prob_N]^R\right]^R.
\end{equation} 
Expanding to second order in $\prob_N$ yields
\begin{equation}
(R(1-\phi)(1+R\phi)-1) \prob_N =  \frac{R(R-1)}{2}(1-\phi)^2 \left(R^2\phi^2 + 3R\phi+1\right) \prob_N^2 + \mathcal{O}\left(\prob_N^3\right).
\end{equation}
In terms of the roots defined in Eq.~(\ref{eq:roots}), we have
\begin{equation}
R^2(\phi_+-\phi)(\phi-\phi_-)\prob_N =  \frac{R(R-1)}{2}(1-\phi)^2 \left(R^2\phi^2 + 3R\phi+1\right) {\prob_N^2}+\mathcal{O}\left(\prob_N^2\right).
\label{eq:probcrit}
\end{equation}
Note that in  Eq.~(\ref{eq:probcrit}), since $\prob_N \geq 0$, the only physical solution for $\phi \geq \phi_+$ is $\prob_N = 0$, whereas $\prob_N > 0$ if $\phi < \phi_+$.
Hence we recover the result $\phi_c = \phi_+$. 
It further follows that
\begin{equation}
1 = \frac{R-1}{2R}(1-\phi)^2\frac{R^2 \phi^2 + 3R\phi+1}{(\phi_c-\phi)(\phi-\phi_-)} \prob_N\left[1 + \mathcal{O}\left(\prob_N\right)+\ldots\right].
\end{equation}
Taking limits as $\phi \to \phi_c^-$ yields
\begin{equation}
\lim_{\phi \to \phi_c^{-}}  \frac{R-1}{2R}(1-\phi)^2\frac{R^2 \phi^2 + 3R\phi+1}{(\phi_c-\phi)(\phi-\phi_-)} \prob_N = 1,
\end{equation}
which implies
\begin{equation}
\lim_{\phi \to \phi_c^{-}} \frac{\prob_N}{(\phi_c-\phi)}  = \frac{2}{R}\sqrt{\frac{R+3}{R+1}} \frac{1}{(R+2)\phi_c+1} \frac{1}{(1-\phi_c)^2},
\end{equation}
and consequently that
\begin{equation}
\prob_N \sim \frac{2}{R}\sqrt{\frac{R+3}{R+1}} \frac{1}{(R+2)\phi_c+1} \frac{1}{(1-\phi_c)^2} (\phi_c - \phi), \quad \phi \to \phi_c^-,
\end{equation}
which suggests a critical exponent $\beta=1$. 
To confirm this exponent, we define the probability of formation of an infinite cluster from a single infected site, $\prob =\phi \prob_C + (1-\phi)\prob_N$. 
Since $\prob_N$ is small in the vicinity of $\phi_c$, we can linearize Eq.~(\ref{eq:Cinfected}),
\begin{equation}
\prob_C = R(1-\phi)\prob_N + \mathcal{O}(\prob_N^2),
\end{equation}
to obtain
\begin{equation}
\prob(\phi) \sim \frac{2}{R}\sqrt{\frac{R+3}{R+1}} \frac{R \phi_c+1}{(R+2)\phi_c+1} \frac{1}{(1-\phi_c)} (\phi_c - \phi), \quad \phi \to \phi_c^-,
\end{equation}
from which the critical exponent $\beta=1$ is immediate.
\section{Exact critical line for digital herd immunity}
\label{AppendixB}
Here, we derive the exact critical line for digital herd immunity using two complementary approaches. 
In App.~\ref{app:meansize}, we obtain recurrence relations for the full probability generating function for the size of infected clusters, which allows us to identify when the mean size of an infected cluster diverges.
In App.~\ref{app:percprob}, we study the processes involved in the cluster growth and determine when the percolation probability of the infected cluster approaches zero.
\subsection{Mean cluster size approach}\label{app:meansize}
First, it is useful to define one probability generating function per type of initial node:
\begin{equation}
B_{\alpha}(\phi,\theta,x) = \sum_{s=1}^\infty \mathbb{P}_{\phi,\theta}(|C| = s | \mathrm{\, initial \, node \, infected \, and \, type\,} \alpha) x^s,
\end{equation}
where $\alpha \in \{CA,CS,NA,NS\}$. 
The probability generating function for cluster sizes, given any type of infected initial node, is then
\begin{align}
B(\phi,\theta,x) &= \sum_{s=1}^\infty \mathbb{P}_{\phi,\theta}(|C| = s | \mathrm{\, initial \, node \, infected})  x^s \nn  \\
&\equiv \sum_{\alpha \in \{CA,CS,NA,NS\}} p_{\alpha}(\phi,\theta) B_{\alpha}(\phi,\theta,x).
\label{eq:Bdefn}
\end{align}
We shall find the exact critical line for $n$-step contact tracing by determining when the mean size of an infected cluster, $\mathbb{E}_{\phi,\theta}(|C|) = \partial_x B(\phi,\theta,x) \big{|}_{x=1}$, diverges.
We first obtain exact recurrence relations for the generating functions $B_{\alpha}$ by enumerating possibilities at a given node.
When the initial infected node is off the contact network, i.e. of type $NA$ or $NS$, we obtain
\begin{equation}
    B_{NA} = xB^{R_0},\;\; B_{NS} = xB^{R_S},
%\label{eq:BN}
\end{equation}
since nodes off the network can infect any other type of node (cf. Eq.~(\ref{eq:BN})). 

When the initial infected node is of type $CS$, any infections in the next generation that are on the network will be detected (cf. Eq.~(\ref{eq:BC})), and we obtain
\begin{equation}
    B_{CS} = x (p_{NA}B_{NA} + p_{NS} B_{NS} + (p_{CA} + p_{CS}) x)^{R_S} \equiv xB_D^{R_S},
\label{eq:Bcs}
\end{equation}
where it is useful to define
\begin{equation}
B_D \equiv B_N + p_C x,\;\;B_N \equiv p_{NA}B_{NA} + p_{NS}B_{NS}, \;\;p_C \equiv p_{CA}+p_{CS}.
\label{eq:Bdefns}
\end{equation}
The analogous result for $B_{CA}$ is rather more involved.
The essential difficulty is that for $n$-step contact tracing, the recurrence relation for $B_{CA}$ involves $n$ generations beyond the initial node, rather than just one.
This is because the possibility arises of multi-generational clusters of $CA$ nodes, that escape detection until they infect a $CS$ node at some later generation $1<m\leq n$. (Put differently, the underlying branching process is not Markovian.)
The simplest way to proceed is to study all the configurations of clusters originating from an initial infected node of type $CA$, and organize the sum according to the generation in which the $CA$ cluster connected to the initial $CA$ node is detected, schematically
\begin{align}
B_{CA} = x\left[\{ \mathrm{no\,\,detection,\,\, gens.\,\,}j \leq n\} + \sum_{j=1}^n \{\mathrm{first\,\,detection\,\,in\,\,gen.\,\,} j\}\right].
\label{eq:schematic}
\end{align}
We first define a function and its composition to recursively ``propagate" the generating function from one generation to the next, after the addition of a $CA$ node:
\begin{equation}
    f(x,y) = (B_N + p_{CA} x y)^{R_0},\;\;\;f^{(j)}(x,y) = \begin{cases} y & j=0 \\ \underbrace{f(x,f(x,\ldots f(x,y) \ldots))}_{j\,\,\mathrm{times}} & j\geq 1 \end{cases}.
\end{equation}
The generating function for the processes without detection in generations $j \leq n$ reads
\begin{equation}
    \{ \mathrm{no\,\,detection,\,\, gens.\,\,}j \leq n\} = f^{(n-1)}(x, g^{(1)}),\;\; g^{(1)} = (B_N + p_{CA}B_{CA})^{R_0}
\label{eq:detectgen}
\end{equation}
where $g^{(1)}$ is the generating function for all processes that do not lead to a $CS$ node in one generation of disease spread.
Similarly the generating function for all processes that end in detection in generation $j$, reads
\begin{eqnarray}
    &\{\mathrm{first\,\,detection\,\,in\,\,gen.\,\,} j\} = (f^{(j)}(x,g^{(2)}) - f^{(j)}(x,g^{(3)})),\nn \\
    &g^{(2)} = (B_N + p_{CS}B_{CS}+p_{CA}xB_D^{R_0})^{R_0},\;\; g^{(3)} = (B_N + p_{CA} x B_D^{R_0})^{R_0},
\label{eq:nodetectgen}
\end{eqnarray}
where $g^{(2)}$ (resp. $g^{(3)}$) are generating functions for all processes that lead to (resp. do not lead to) creation of a $CS$ node in generation $j$, and the subtraction ensures that only processes that give rise to at least one $CS$ node in generation $j$ are included.
Using Eqs.~(\ref{eq:schematic}), (\ref{eq:detectgen}), and (\ref{eq:nodetectgen}), we find that
\begin{align}
B_{CA} = x\left[ f^{(n-1)}(x,g^{(1)}) + \sum_{j=0}^{n-1} (f^{(j)}(x,g^{(2)}) - f^{(j)}(x,g^{(3)})) \right].
\label{eq:BCAexpr}
\end{align}
It can be verified that $B_{CA}$ is correctly normalized, i.e. that $B_{CA}\big{|}_{x=1} = 1$.
To compute the mean cluster size, we first compute derivatives of the generating functions $B_\alpha$. 
Noting that $B_\alpha\big{|}_{x=1}= 1$ and using Eqs.~(\ref{eq:BN}) and (\ref{eq:Bcs}), we obtain
\begin{eqnarray}
    &\partial_x B_{NA}\big{|}_{x = 1} = 1 + R_0 \partial_x B\big{|}_{x = 1},\;\; \partial_x B_{NS}\big{|}_{x = 1} = 1 + R_S \partial_x B\big{|}_{x = 1},\nn \\
    &\partial_x B_{CS}\big{|}_{x = 1} = 1 + R_S \partial_x B_D\big{|}_{x = 1} = 1 + R_S (p_C + p_{NA} \partial_x B_{NA}\big{|}_{x = 1} + p_{NS} \partial_x B_{NS}\big{|}_{x = 1}).
\label{eq:derivs}
\end{eqnarray}
Eliminating all variables other than $\partial_x B \big{|}_{x=1}$, we can write the derivative of $B_{CA}$ in the form
\begin{equation}
\partial_{x} B_{CA}\big{|}_{x=1} = E_n + F_n \partial_x B \big{|}_{x=1},
\label{eq:BCAderiv}
\end{equation}
and in terms of these coefficients $\{E_n, F_n\}$, the mean cluster size reads
\begin{align}
\mathbb{E}_{\phi,\theta}(|C|) = \partial_x B \big{|}_{x=1} = \frac{1-p_{CA}+p_{CS}R_S + p_{CA}E_n}{1-(R_N + p_{CS}R_SR_N + p_{CA} F_n)},
\end{align}
where is is useful to define
\begin{equation}
R_N = p_{NA} R_0 + p_{NS}R_S, \quad R_C = p_{CA} R_0 + p_{CS}R_S.
\label{eq:RNRCdefn}
\end{equation}
It is clear that the mean cluster size diverges when
\begin{equation}
R_N + p_{CS}R_SR_N + p_{CA} F_n = 1.
\end{equation}
It remains to compute $F_n$, as defined in Eq.~(\ref{eq:BCAderiv}). To this end, let us introduce functions
\begin{align}
c^{(i)}_j = \partial_x f^{(j)}(x, y)\big{|}_{x=1, y = g^{(i)}(1)}
\end{align}
of $\phi$ and $\theta$, with arguments other than $x$ suppressed. By the chain rule, these satisfy the recurrence relations
\begin{align}
\nonumber c^{(i)}_{j+1} &= \partial_x f(x,f^{(j)}(x, y)) \big{|}_{x=1, y = g^{(i)}(1)} \\
\nn &= \partial_x f(1, f^{(j)}(1,g^{(i)}(1))) + \partial_y f(1,f^{(j)}(1,g^{(i)}(1))) \partial_x f^{(j)}(1,g^{(i)}(1)))\\
&= R_0 (\partial_x B_N\big{|}_{x=1} + p_{CA}a^{(i)}_j)(p_N + p_{CA} a^{(i)}_j)^{R_0-1} + R_0 p_{CA} (B_N\big{|}_{x=1} + p_{CA} a^{(i)}_j)^{R_0-1} c^{(i)}_j,
\label{eq:chainrule}
\end{align}
where we defined $a^{(i)}_j = f^{(j)}(1,g^{(i)}(1))$. Since we are only concerned with the coefficient of $\partial_x B\big{|}_{x=1}$ in $\partial_x B_{CA}\big{|}_{x=1}$, let us write $c^{(i)}_j = d^{(i)}_j + b^{(i)}_j \partial_x B\big{|}_{x=1}$, as in Eq.~(\ref{eq:BCAderiv}). Upon making this substitution in Eq.~(\ref{eq:chainrule}), we obtain the recurrence
\begin{eqnarray}
b^{(i)}_{j+1} = R_0 (p_N + p_{CA} a^{(i)}_j)^{R_0-1} (R_N + p_{CA}b_j^{(i)})
\end{eqnarray}
for the terms $b^{(i)}_j$ of interest. Combining the above expressions, we find
\begin{equation}
F_n = b^{(1)}_{n-1}+\sum_{j=0}^{n-1}\left(b_j^{(2)}-b_j^{(3)}\right),
\end{equation}
where the $b^{(i)}_j$ are defined recursively for $j>0$ via
\begin{align}
b^{(i)}_{j+1} &= R_0(p_N + p_{CA}a^{(i)}_j)^{R_0-1}(R_N + p_{CA}b^{(i)}_j),\nn \\
a^{(i)}_{j+1} &= (p_{N}+p_{CA}a^{(i)}_j)^{R_0},
\end{align}
with
\begin{align}
b^{(1)}_0 &= R_0(1-p_{CS})^{R_0-1}(1-p_{CS}R_SR_N), \nn \\
b^{(2)}_0 &= R_0 R_N (1+R_C), \nn \\
b^{(3)}_0 &= R_0 R_N(1+p_{CA}R_0)(1-p_{CS})^{R_0-1},
\end{align}
and
\begin{equation}
a_0^{(1)} = a_0^{(3)} = (1-p_{CS})^{R_0}, \quad a_0^{(2)} = 1.
\end{equation}
The exact critical line for $n$-step contact tracing is thus given by the implicit equation
\begin{align}
R_n(\phi,\theta) = R_N + p_{CS}R_SR_N + p_{CA}\left[b^{(1)}_{n-1}+\sum_{j=0}^{n-1}{\left(b_j^{(2)}-b_j^{(3)}\right)}\right] = 1
\label{eq:exactcrit}
\end{align}
for $\phi$ and $\theta$.
\subsection{Percolation probability approach}\label{app:percprob}
We now outline a procedure to obtain the critical line using the percolation probability of an infected initial node.
The probability of formation of an infinite cluster is given by
\begin{equation}
    \prob \equiv \sumal{\alpha \in \{CA, CS, NA, NS\}}{}{p_\alpha \prob_\alpha},
\label{eq:pexpr}
\end{equation}
where $\prob_\alpha$ is the probability of formation of an infinite cluster starting from a node of type $\alpha$. 
Since nodes off the network ($NS$ and $NA$) can infect any type of node, we obtain the recurrence relations
\begin{equation}
    \prob_{NS} = 1 - (1 - \prob)^{R_S},\;\;\; \prob_{NA} = 1 - (1 - \prob)^{R_0}.
\label{eq:probNrec}
\end{equation}
Further, since the $CS$ node can only infect anyone outside the network, we obtain
\begin{equation}
    \prob_{CS} = 1 - (1 - \prob_N)^{R_S}, \;\; \prob_N \equiv p_{NS} \prob_{NS} + p_{NA} \prob_{NA}.
\label{eq:probCSrec}
\end{equation}
Similar to the case of the generating function in the previous section, obtaining the recurrence relation for $\prob_{CA}$ is more involved.
We proceed by enumerating the minimum number of processes such that the expression for the formation of an infinite cluster $\prob_{CA}$ can be expressed in terms of the $\prob_\alpha$'s.
For depth $n$ contact tracing, this yields a term that schematically reads
\begin{equation}
    \prob_{CA} = \{\textrm{no detection, gens. $j\leq n$\}} + \sumal{j = 1}{n}{\{\textrm{first detection in gen. j}\}}.
\label{eq:pcaprocess}
\end{equation}
To determine the critical line, it is sufficient to linearize the recurrence relations for $\{\prob_\alpha\}$ similar to the calculation in App.~\ref{AppendixA}.
We thus obtain
\begin{equation}
    \prob_{NS} = R_S \prob + \mathcal{O}(\prob^2),\;\; \prob_{NA} = R_0 \prob + \mathcal{O}(\prob^2),\;\; \prob_{CS} = R_S \prob_{N} + \mathcal{O}(\prob^2) = R_S R_N \prob + \mathcal{O}(\prob^2),
\label{eq:pslinear}
\end{equation}
where $R_N$ is defined in Eq.~(\ref{eq:RNRCdefn}). 
To count the different kind of processes that enter Eq.~(\ref{eq:pcaprocess}), we divide all processes of $j$ generations into three types, which we denote as follows. 
\begin{align}
    \mathcal{C}^{(j)} &: \{\textrm{Processes with at least one $CA$ in generation $j$}\}\nn \\
    \mathcal{E}^{(j)} &: \{\textrm{Processes with no $CA$ or $CS$ in generation $j$}\}\nn \\
    \mathcal{T}^{(j)} &: \{\textrm{Processes with at least one $CS$ in generation $j$}\}.
\label{eq:types}
\end{align}
\begin{figure}
    \centering
    \includegraphics[width=\textwidth,trim=0cm 13.5cm 0cm 8.5cm]{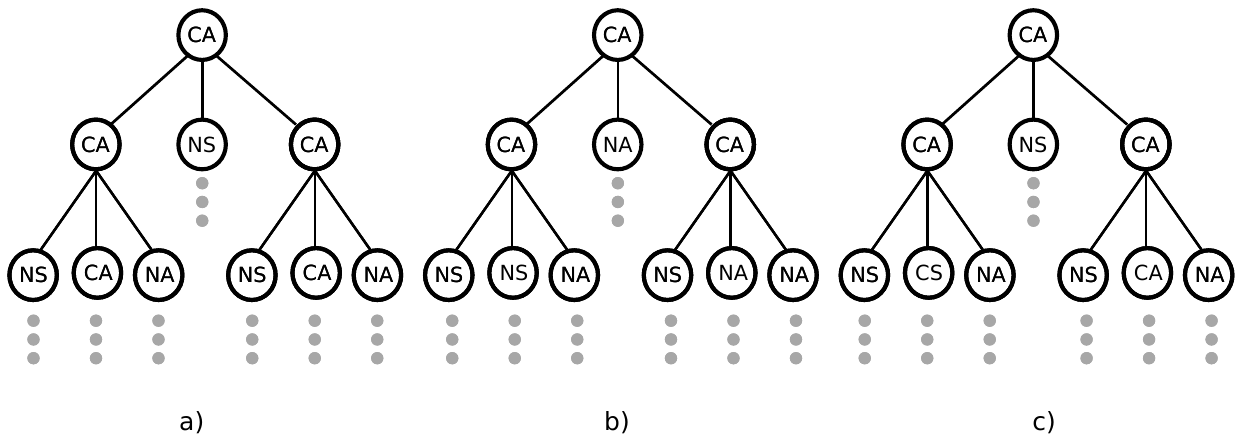}
    \caption{Examples of processes that appear in the recursive expression for $\prob_{CA}$. Processes shown in (a), (b), and (c) belong to the sets $\mathcal{C}^{(2)}$, $\mathcal{E}^{(2)}$, and $\mathcal{T}^{(2)}$ respectively (see Eq.~(\ref{eq:types})). $\prob_{CA}$ can be expressed in terms of the probabilities of formation of infinite clusters starting from the nodes on the edge of the processes.}
    \label{fig:percprobtree}
\end{figure}
Note that processes in $\mathcal{C}^{(j)}$ and $\mathcal{E}^{(j)}$ lead to no detection in generation $j$ and $\mathcal{T}^{(j)}$ leads to a detection in generation $j$ that activates the contact network. 
Examples of these processes for $j = 2$ are shown in Fig.~\ref{fig:percprobtree}.
To linear order in $p$, we find that the terms in $\prob_{CA}$ in Eq.~(\ref{eq:pcaprocess}) can be expressed as
\begin{eqnarray}
    &&\{\textrm{no detection, gens. $j\leq n$\}} = \frac{1}{p_{CA}}\left(\sumal{\tau \in \mathcal{C}^{(n)}}{}{P_\tau \sumal{\alpha}{}{n_\alpha(\tau) \prob_\alpha}} + \sumal{j = 1}{n}{\sumal{\tau \in \mathcal{E}^{(j)}}{}{P_\tau \sumal{\alpha}{}{n_\alpha(\tau) \prob_\alpha}}}\right) + \mathcal{O}(\prob^2) \nn \\
    &&\{\textrm{first detection in gen. j}\} = \frac{1}{p_{CA}}\left(\sumal{\tau \in \mathcal{T}^{(j)}}{}{P_\tau \left(n_{CA}(\tau) \prob^{(D)}_{CA} + \sumal{\alpha \in \{CS, NS, NA\}}{}{n_{\alpha}(\tau)\prob_\alpha}\right)}\right) + \mathcal{O}(\prob^2),
\label{eq:pcalin}
\end{eqnarray}
where $\tau$ runs over all the processes in the sets described in Eq.~(\ref{eq:types}), $\alpha$ is summed over all types of nodes unless other wise stated, and we have defined $n_\alpha(\tau)$, $P_\tau$, and $\prob^{(D)}_{CA}$ as follows.
$n_\alpha(\tau)$ denotes the number of nodes of the type $\alpha$ on the edge of a process $\tau$.
$P_\tau$ is the probability of having a process $\tau$, i.e. the product of the individual probabilities $\{p_\alpha\}$ of all the nodes in the process including the root $CA$ node, and in Eq.~(\ref{eq:pcalin}) we divide by a factor of $p_{CA}$ in order to determine the probabilities of the processes given that the root is of type $CA$.
$\prob^{(D)}_{CA}$ is the probability of formation of an infinite cluster due a $CA$ that has been detected due to a $CS$ node in the same generation, which reads
\begin{equation}
    \prob^{(D)}_{CA} = 1 - (1 - \prob_N)^{R_0} = R_0 \prob_N + \mathcal{O}(\prob^2) = R_0 R_N \prob + \mathcal{O}(\prob^2),
\label{eq:detectedCA}
\end{equation}
where we have used Eq.~(\ref{eq:pslinear}). 
To evaluate the terms in Eq.~(\ref{eq:pcalin}), we define generating functions corresponding to each of the processes in Eq.~(\ref{eq:types}) as 
\begin{equation}
    C^{(j)}(\{z_\alpha\}) \equiv \sumal{\tau \in \mathcal{C}^{(j)}}{}{P_\tau\prodal{\alpha}{}{ z_\alpha^{n_\alpha(\tau)}}},\;\;\;E^{(j)}(\{z_\alpha\}) \equiv \sumal{\tau \in \mathcal{E}^{(j)}}{}{P_\tau \prodal{\alpha}{}{z_\alpha^{n_\alpha(\tau)}}},\;\;\;T^{(j)}(\{z_\alpha\}) \equiv \sumal{\tau \in \mathcal{T}^{(j)}}{}{P_\tau \prodal{\alpha}{}{z_\alpha^{n_\alpha(\tau)}}}.
\end{equation} 
These generating functions can be enumerated recursively as 
\begin{align}
    C^{(n)} &= p_{CA}\left(\left(C^{(n-1)} + \sumal{j = 0}{n-1}{E^{(j)}}\right)^{R_0} - \left(\sumal{j = 0}{n-1}{E^{(j)}}\right)^{R_0}\right)\nn \\
    E^{(n)} &= p_{CA}\left(\left(\sumal{j = 0}{n-1}{E^{(j)}}\right)^{R_0} - \left(\sumal{j = 0}{n-2}{E^{(j)}}\right)^{R_0}\right) \nn \\
    T^{(n)} &= p_{CA}\left(\left(C^{(n-1)} + T^{(n-1)} + \sumal{j = 0}{n-1}{E^{(j)}}\right)^{R_0} - \left(C^{(n-1)} + \sumal{j = 0}{n-1}{E^{(j)}}\right)^{R_0}\right)
\end{align}
with
\begin{equation}
    C^{(0)} = p_{CA} z_{CA},\;\; E^{(0)} = p_{NS} z_{NS} + p_{NA} z_{NA},\;\; T^{(0)} = p_{CS} z_{CS}. 
\end{equation}
In terms of these generating functions, Eq.~(\ref{eq:pcalin}) reads 
\begin{eqnarray}
    &&\{\textrm{no detection, gens. $j\leq n$\}} = \frac{1}{p_{CA}}\sumal{\alpha}{}{\prob_\alpha\ \partial_{z_\alpha} \left(C^{(n)} + \sumal{j = 1}{n}{E^{(j)}}\right)}\big{|}_{\{z_\alpha = 1\}} + \mathcal{O}(\prob^2) \nn \\
    &&\{\textrm{first detection in gen. j}\} = \frac{1}{p_{CA}}\left(\prob^{(D)}_{CA}\ \partial_{z_{CA}} T^{(j)} + \sumal{\alpha \in \{CS, NS, NA\}}{}{\prob_\alpha\ \partial_{z_\alpha} T^{(j)}}\right)\big{|}_{\{z_\alpha = 1\}} + \mathcal{O}(\prob^2).
\label{eq:pcalin2}
\end{eqnarray}
Using Eqs.~(\ref{eq:pexpr}), (\ref{eq:pslinear}), (\ref{eq:detectedCA}), and (\ref{eq:pcalin2}), we finally obtain an expression for $\prob$ of the form
\begin{equation}
    \prob = R_n(\phi, \theta) \prob + \mathcal{O}(\prob^2), 
\end{equation}
using which we identify the critical line to be
\begin{equation}
    R_n(\phi_c, \theta_c) = 1. 
\end{equation}
While we are not able to derive a more explicit expression for $R_n(\phi, \theta)$ using this approach, we have verified using \textsc{Mathematica} that it yields the same critical line of Eq.~(\ref{eq:exactcrit}) for several values of $n$.
That is, we find that
\begin{equation}
R_n(\phi,\theta) = R_N + p_{CS}R_S R_N + p_{CA}\left[b^{(1)}_{n-1}+\sum_{j=0}^{n-1}\left(b_j^{(2)}-b_j^{(3)}\right)\right].
\label{eq:rndefn}
\end{equation}
\subsection{Critical exponents}
\begin{figure}[h]
    \centering
    \includegraphics[width=0.45\linewidth]{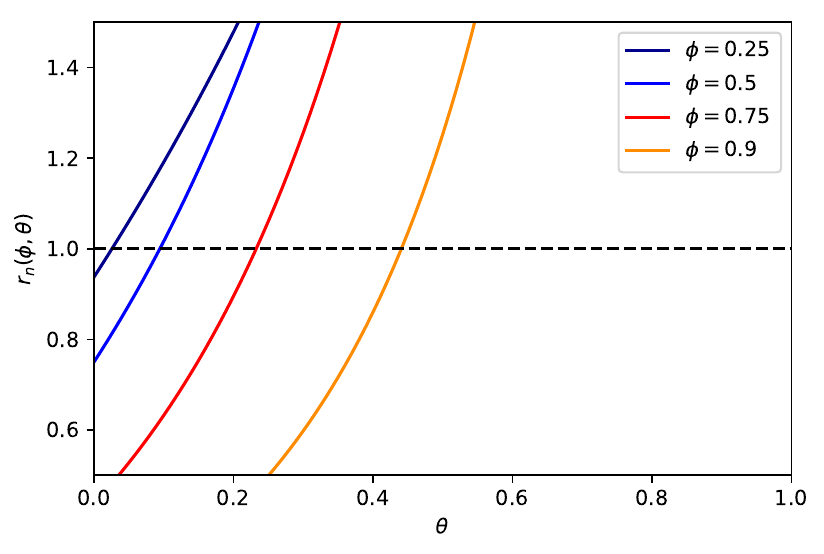}
    \includegraphics[width=0.45\linewidth]{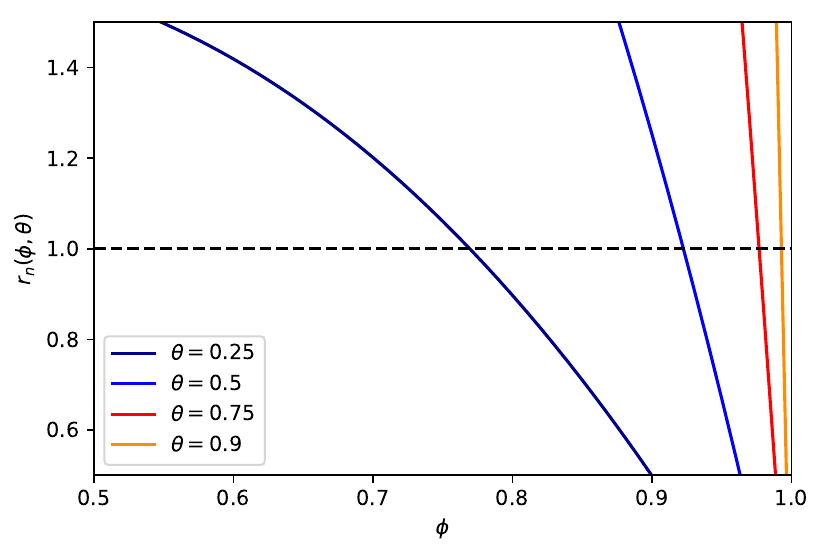}
    \caption{Behaviour of $R_n(\phi,\theta)$ near the critical line $R_n(\phi,\theta)=1$ for various fixed $\phi$ and continuously varying $\theta$, and vice versa. It is clear that $R_n(\phi,\theta)$ behaves linearly in $\theta$ and $\phi$ near the critical line $R_n=1$. Results are shown for a contact-tracing depth $n=100$ and parameter values $R_0=3$, $R_S=1$; the summation was checked to be well-converged already for $n=10$.}
    \label{FigApp}
\end{figure}

Following the derivation of critical exponents in App.~\ref{AppendixA}, it is clear that for any finite $n$, the critical exponents $\beta$ and $\gamma$ are controlled by the behaviour of the function
near the critical line $R_n (\theta, \phi) = 1$.
Unfortunately, it does not seem possible to obtain $R_n(\phi,\theta)$ in closed form. 
However, since $R_n (\phi, \theta)$ is analytic for any finite $n$, the critical behaviour on the transition line is expected to be determined by the leading terms in its Taylor expansion, which are linear in $\phi$ for given $\theta$ and vice versa. 
Numerically, we find that this is indeed the case; see Fig.~\ref{FigApp}. The numerical data suggests that $\gamma=\beta=1$ everywhere except on the line $\phi=1$, whose critical behaviour is discussed below.

\section{Critical behaviour along the line $\phi=1$}
\label{AppendixC}
Here, we study the critical behaviour along the line $\phi=1$. 
In the limit of infinite tracing depth, we find that there is a discontinuous phase transition at $\theta_c=1$. However, for any finite contact-tracing depth $n$, the critical point $\theta_c<1$, and the transition shows the same critical exponents as mean-field percolation.
\subsection{Infinite tracing depth}
When the tracing depth is infinite, there is a discontinuous transition as $\theta \to 1^-$, in the sense that the ``order parameter'', i.e. the probability $\rho(\theta)$ of formation of an infinite infected cluster, jumps discontinuously from $0$ to $1$ at $\theta_c=1$.
To see this, note that when $\phi=1$, the evolution of an asymptomatic cluster from an asymptomatic source is described by the following branching process
\begin{align}
&X^A_0 = 1,\;\;\; X^A_1 = Z^A_{0,1}, \nn \\
&X^A_{n+1} = \left[Z^A_{n,1}+Z^A_{n,2}+\ldots+Z^A_{n,X^A_n}\right]\mathbbm{1}_{X_n^A=R_0^{n}},
\end{align}
where $X_n^A$ denotes the total number of new asymptomatic infections in generation $n$, the indicator functions $\mathbbm{1}_{X_n^A=R_0^{n}}$ reflect the fact that a single symptomatic case will terminate the branching process, and
\begin{equation}
Z^A_{i,j} \sim \mathrm{Bin}(R_0,\theta) \quad \mathrm{i.i.d.}
\end{equation} 
The probability generating function of $X^A_n$ then satisfies the recurrence relation
\begin{align}
\nonumber F_{n+1}(y) &= \mathbb{E}[y^{X_{n+1}^A}] \\
\nonumber &= \mathbb{E}\left[\mathbb{E}[y^{X_{n+1}^A} | X_{n}^A]\right] \\
\nonumber &= 1-\mathbb{P}(X_{n}^A=R_0^{n}) + \mathbb{P}(X_{n}^A=R_0^{n}) \mathbb{E}\left[y^{\sum_{j=1}^{R_0^n} Z_{n,j}^A}\right]  \\
&= 1-\mathbb{P}(X_{n}^A=R_0^{n}) + \mathbb{P}(X_{n}^A=R_0^{n}) f(y)^{R_0^{n}},
\end{align}
where $f(y)$ is the probability generating function for the descendants of a single asymptomatic node,
\begin{equation}
f(y) = (\theta y + (1-\theta))^{R_0}.
\end{equation}
Using these results, it can be shown by induction that
\begin{equation}
F_n(y) = 1 - \theta^{R_0+R_0^2+\ldots+R_0^{n-1}} + \theta^{R_0+R_0^2+\ldots+R_0^{n-1}} f(y)^{R_0^{n-1}}.
\end{equation}
The probability that the branching process is extinct in generation $n$ is then
\begin{equation}
F_n(0) = 1 - \theta^{R_0+R_0^2+\ldots+R_0^{n-1}} + \theta^{R_0+R_0^2+\ldots+R_0^{n-1}} (1-\theta)^{R_0^n}.
\end{equation}
The extinction probability is thus
\begin{equation}
q = \lim_{n\to \infty} F_n(0) = \begin{cases} 1 & \theta< 1 \\ 0 & \theta = 1\end{cases}.
\end{equation}
It follows that the critical point occurs at $\theta = \theta_c = 1$, and that the probability of formation of an infinite cluster, which is usually regarded as the order parameter for percolation transitions, behaves like
\begin{equation}
\rho = 1-q =\begin{cases} 0 & \theta< 1 \\ 1 & \theta = 1\end{cases}.
\end{equation}
Such discontinuous behaviour of the order parameter indicates that the transition has a first-order character. For example, at $\theta=1$ all asymptomatic clusters are infinite and the critical exponent $\delta$ is not even defined. However, several other critical exponents are defined, a scenario reminiscent of one-dimensional site percolation, which also has a critical probability $p_c=1$ and a mixture of continuous and discontinuous behaviour as $p \to p_c^{-}$.
\subsection{Finite tracing depth}
For any finite tracing depth $n$, the exact critical point along the line $\phi=1$ is found to lie at
\begin{equation}
\theta_c(n,R_0) = \left(\frac{1}{R_0^n}\right)^{\frac{1}{R_0+R_0^2+\ldots + R_0^n}},
\label{eq:phi1crit}
\end{equation}
with mean-field critical exponents, $\gamma=\beta=1$.
While this can be obtained from the results of App.~\ref{AppendixB}, we provide an intuitive explanation here.
When $\phi = 1$, all nodes are on the contact network, and hence only two types need to be considered: symptomatic ($S$) and asymptomatic ($A$), which occur with probabilities $(1 - \theta)$ and $\theta$ respectively.
Since any infinite cluster must consist entirely of $A$ nodes, we can directly obtain the recurrence relation for the probability $\rho$ of formation of an infinite cluster. 
For contact tracing with recursive depth $n$, the only event that does not rule out the existence of an infinite cluster is the formation of an $n$-generation tree in which every node is asymptomatic. This occurs with probability $\theta^{N_n}$, where
\begin{equation}
    N_n = R_0 + R_0^2 + \cdots + R_0^n,
\label{eq:totnode}
\end{equation}
where $N_n$ is the total number of nodes in such a tree excluding the root node. 

Given such a tree, infinite clusters can originate from any of the $R_0^n$ asymptomatic leaves in generation $n$.
This yields the following recurrence relation for $\rho$:
\begin{equation}
    \rho = \theta^{N_n} \left(1 - (1 - \rho)^{R_0^n}\right).
\label{eq:pinfrec}
\end{equation}
Expanding Eq.~(\ref{eq:pinfrec}) to second order in $\rho$, it is straightforward to derive Eq.~(\ref{eq:phi1crit}) and that
\begin{equation}
    \rho \sim \frac{2 N_n}{(R_0^n - 1)\theta_c}(\theta - \theta_c). 
\end{equation}
Thus, $\beta = 1$ for any finite $n$. A similar argument yields the mean cluster size
\begin{equation}
\mathbb{E}_{\theta}(|C|) \propto \frac{1}{1-R_0^n\theta^{N_n}},\quad \theta < \theta_c,
\end{equation}
which has the same critical point as the percolation probability, and yields a critical exponent $\gamma=1$, since
\begin{equation}
\mathbb{E}_{\theta}(|C|) \sim \frac{1}{(\theta_c-\theta)}, \quad \theta \to \theta_c^{-}.
\end{equation}

\section{Mean-field-like estimate for the critical line}
\label{AppendixD}
In this section, we derive a simple, approximate formula for $R = R_\infty(\phi,\theta)$, based on linear interpolation and a ``mean-field'' assumption, whereby inter-generational correlations are neglected.
It is first helpful to label the possible types of infected individual by $\alpha \in \{CA, CS, NA, NS\}$, and note that susceptible individuals of each type occur with the independent probabilities given in Table \ref{Tab1}:
\begin{equation}
\label{Tab1}
\begin{tabular}{c|c}
Type $\alpha$ & $p_{\alpha}$ \\
    \hline
    CA & $\phi\theta$ \\
    CS & $\phi(1-\theta)$\\
    NA & $(1-\phi)\theta$ \\
    NS & $(1-\phi)(1-\theta)$
\end{tabular}
\end{equation}
Now suppose that there are no correlations between generations. Then the effective reproduction number is simply an average over the possible types of node:
\begin{equation}
\label{MF}
    R = \sum_{\alpha \in \{CA, CS, NA, NS\}} p_{\alpha} R_{\alpha}.
\end{equation}
Here, the probabilities $p_{\alpha}$ are given as in Table \ref{Tab1}, while  $R_{NA} = R_0$, $R_{NS}=R_S$, and $R_{CS} = R_S(1-\phi)$, corresponding to the average number of live nodes generated by a symptomatic individual on the contact network. However, $R_{CA}$ is essentially undetermined in this approach, since discarding correlations in time also discards contact tracing, to which the effective value of $R_{CA}$ is highly sensitive. We will therefore treat $R_{CA}$ as a variational parameter, to be estimated self-consistently. In the ``best'' case, asymptomatic transmission within the network is completely suppressed, and $R_{CA} = (1-\phi)R_0$. In the ``worst'' case, asymptomatic transmission within the network is not suppressed at all, and $R_{CA} = R_0$. A simple way to proceed is to solve for the unique linear interpolation between these cases that passes through the known endpoint $(\phi,\theta) = (1,1)$ of the non-perturbative critical line (see Appendix \ref{AppendixC}). The resulting approximation to $R$ is given by
\begin{equation}
\label{ApproxR}
R \approx R_S (1 - \theta)(1-\phi^2) + R_0 \theta\left(1 - \left(1 - \frac{1}{R_0}\right) \phi^2\right).
\end{equation}
\section{Estimating the number of tests required for successful contact tracing}
\label{AppendixE}
The number of tests per symptomatic infection required to successfully implement the digital contact tracing scheme described in this paper depends on the basic reproduction number $R_0$, the fraction of asymptomatic cases $\theta$, the recursive depth of contact tracing $n$, the typical number of contacts per individual $N$ and the number of new cases $N_C$ per day (this is the true number of cases, that includes undetected, asymptomatic infections). In this Appendix, we estimate the number of tests required to implement single-step contact tracing ($n=1$); our calculation is easily generalized to larger recursive depths $n>1$.

First suppose that $R_0< 1/\theta$. Then isolating $(1-\theta)N_C$ symptomatic cases and tracing their contacts during their possible pre-symptomatic contagious period, for a total of $N_T = N (1-\theta)N_C$ tests per day, leaves $\theta N_C$ asymptomatic cases per day that can spread the disease. In the next generation, these give rise to $R_0\theta N_C$ new infections. The effective reproduction $R$ is then given by
\begin{equation}
R = \frac{\mathrm{new \, cases}}{\mathrm{old \, cases}} = \frac{R_0 \theta N_C}{N_C} = R_0\theta < 1.
\end{equation}
Thus epidemic spread is prevented within a handful of disease generations.

Next suppose that $R_0>1/\theta$. Again, we start by testing the symptomatic cases and the contacts they encountered during their pre-symptomatic contagious period, leading to $N_T = N (1-\theta)N_C$ tests per day. During the second generation of infections, this cohort yields $(1-\theta) R_0\theta N_C$ hitherto undetected symptomatic cases per day, all of which arise from asymptomatic sources. For these cases, there are $R_0$ distinct possibilities that need to be considered. The probability that a given asymptomatic source generates $k$ symptomatic infections is given by
\begin{equation}
\label{eq:eventk}
\mathbb{P}(\, k \mathrm{\,\, symptomatic \,\,cases}\,|\mathrm{\,asymptomatic\,\,source}\,) =  \begin{pmatrix} R_0 \\ k \end{pmatrix} (1-\theta)^k \theta^{R_0-k}, 
\end{equation}
and the average number of tests per symptomatic case is $(R_0+k+1)\times N/k$, where $k=1,2,\ldots,R_0$, which can be understood as follows. For each symptomatic infection, their contacts from the pre-symptomatic period as well as the period when they were likely infected need to be tested, yielding a total of $k\times 2N$ tests. Once the asymptomatic source is found, we must additionally trace their $N$ contacts, to find $R_0-k$ asymptomatic infections in the current generation, and finally test \emph{their} $N$ contacts to prevent future transmission. Thus the total number of contacts traced in the event Eq. \eqref{eq:eventk} is $(2k + 1 + R_0-k)\times N = (R_0+k+1)\times N$. Putting this all together, the average number of contacts that need to be tested per symptomatic case is
\begin{equation}
N_{\mathrm{eff}} = \left[ \sum_{k=1}^{R_0} \begin{pmatrix} R_0 \\ k \end{pmatrix} \left(\frac{R_0+k+1}{k}\right) (1-\theta)^k \theta^{R_0-k}  \right]N,
\end{equation}
in terms of which the required daily testing rate is
\begin{equation}
N_T = N_{\mathrm{eff}}\theta R_0 (1-\theta)N_C.
\end{equation}
As an illustrative example, with $R_0=3, \theta=1/2$, $N_{\mathrm{eff}}=3.3NN_C$, and assuming around 10 significant contacts per person and around $50,000$ cases per day, $N_T\approx 1.2$ million tests per day.

The effective $R$ achievable in this manner is
\begin{equation}
R=R_0\theta^{R_0}.
\end{equation}
Notice that $R<1$ for $\theta < \theta_c$, where the critical value of asymptomatic transmission is given by
\begin{equation}
\theta_c = (1/R_0)^{1/R_0}.
\end{equation}
For $R_0=3$, this is $\theta_c \approx 0.69$. For higher rates of asymptomatic transmission, higher orders of recursive contact tracing are necessary to prevent an epidemic, with a concomitant increase in the number of daily tests.

\end{document}